\documentclass[twocolumn,aps,prl,superscriptaddress,preprintnumbers,amsmath,amssymb,floatfix]{revtex4-2}

\usepackage{graphicx}
\usepackage{booktabs}
\usepackage{xspace}
\usepackage[dvipsnames]{xcolor}
\usepackage{amsmath}
\usepackage{svg}
\usepackage{bm}
\usepackage{hyperref}
\definecolor{rkcolor}{RGB}{200,30,30}

\newcommand{\cc}[1]{{\color{violet}#1}}
\DeclareRobustCommand{\mg}[1]{{\color{orange}#1}}
\usepackage[normalem]{ulem}
\DeclareRobustCommand{\mgdel}[1]{{\color{orange}\sout{#1}}}

\graphicspath{{./figures/}}

\begin{document}
	
	\title{Giant Raman Optical Activity in Chiral and Polar Ni$_3$TeO$_6$}
	\author{Philipp~Stein}
	\affiliation{Institute of Physics II, University of Cologne, D-50937 Cologne, Germany}
	\author{Ravi Kaushik}
	\affiliation{Quantum Materials theory, 16162 Genova, Italy}
	\author{Evgenia V. Komleva}
	\affiliation{M. N. Mikheev Institute of Metal Physics UB RAS, Ekaterinburg 620137, Russia}
		\author{Thomas C. Koethe}
	\affiliation{Institute of Physics II, University of Cologne, D-50937 Cologne, Germany}
	\author{Vladimir Tsurkan}
	\affiliation{Experimental Physics V, University of Augsburg, D-86159 Augsburg, Germany}
	\author{Istv\'an K\'ezsm\'arki}
	\affiliation{Experimental Physics V, University of Augsburg, D-86159 Augsburg, Germany}
	\author{Markus Gr\"uninger}
	\affiliation{Institute of Physics II, University of Cologne, D-50937 Cologne, Germany}
	\author{Sergey V. Streltsov}
	\affiliation{M. N. Mikheev Institute of Metal Physics UB RAS, Ekaterinburg 620137, Russia}
	\author{Sergey Artyukhin}
	\affiliation{Quantum Materials theory, 16162 Genova, Italy}
	\author{Paul H. M. van Loosdrecht}
	\affiliation{Institute of Physics II, University of Cologne, D-50937 Cologne, Germany}
	
	\begin{abstract}
Raman optical activity (ROA) measures the difference in Raman scattering intensity for right- and left-circularly polarized light and reflects the chirality of excitations. In solids, ROA is commonly very weak. For the phonons of chiral and polar Ni$_3$TeO$_6$, we find a giant ROA with magnitude 100 times higher than commonly reported for molecules. 
The systematic analysis based on phonon selection rules allows the chiral and achiral scattering channels to be separated. First-principles calculations of the Raman tensor at finite phonon momentum relate the effect to the phonon modulation of the band-geometric and band-dispersion contributions
to optical gyrotropy. This establishes Raman optical activity as a probe
of electron--phonon modulation of the band geometry in an insulator.
	\end{abstract}
	
	\maketitle
	
\section{Introduction}
Chirality, the breaking of all inversion symmetries of static objects or dynamic motions, is one of the fundamental concepts in nature. Its relevance encompasses all fields of science, from the handedness of electroweak interaction to the function of biomolecules in life science. Similarly, it also plays a remarkable role in a plethora of phenomena in condensed matter physics, from the basic, long-known concept of optical activity to more recent discoveries like Klein tunneling of chiral fermions in graphene \cite{Intr_Chirality_1} and orbital momentum monopoles in chiral systems for use in orbitronics \cite{Intr_Chirality_2}.
	Chiral bosons, or specifically chiral phonons, are a much more recent discovery \cite{Intr_ChiralPhononDiscovery}. The term implies a unidirectional circular motion of atoms. They are expected to have a strong impact on the physical properties of materials \cite{Intr_Chirality_3,Intr_Chirality_4,Intr_Chirality_5,Intr_Chirality_6}, specifically applications such as controllable phononic heat transport \cite{Intr_Chirality_7,Intr_Chirality_8,Intr_Chirality_9}. They are also promising as robust information carriers for solid state quantum information applications \cite{Intr_Chirality_10}, specifically because they can be selectively manipulated by coupling with circularly polarized light \cite{Intr_Chirality_8,Intr_Chirality_11}. Raman scattering has recently been rediscovered \cite{Intr_Chirality_8,Intr_Raman_1,Intr_Raman_2,Intr_Raman_3} as a detection method for chiral systems by utilizing the phenomenon of linear wave-vector shifts that lead to energy splitting of degenerate phonon modes close to the $\Gamma$ point \cite{Intr_Raman_4,Intr_Raman_5,Intr_Raman_6,Intr_Raman_7,Intr_Raman_8,Intr_Raman_9}. However, the small momentum transfer from visible light to the material causes the effect to be extremely weak. Using higher energy photons is one way to circumvent this problem \cite{Intr_RIXS}. Another way to detect chirality of vibrations that has been long established in spectroscopy of chiral molecules \cite{Intr_ROA_1,Intr_ROA_2,Intr_ROA_3,Intr_ROA_4} could be Raman optical activity (ROA), which instead of an energy splitting measures the difference in scattering intensity for incident left- or right-circularly polarized light (L/R). This effect, first described by Barron and Buckingham in 1971 \cite{Intr_ROA_1}, originates from the finite influence of magnetic dipole- and electric quadrupole transition moments in chiral systems. Besides the conventionally considered interaction in Raman scattering, which assumes electric dipole transitions for both the absorption and emission part of the scattering process, this creates additional, chiral scattering terms in the Raman intensity:
	\begin{equation}
		\centering
		I_{\mathrm{Raman}}  \propto  I_{pp}  + I_{pm}  + I_{p\Theta}  + \mathrm{higher\ orders}.
	\end{equation}
	Here, $I_{pp}$ is the usual achiral contribution due to purely electric dipole processes, while $I_{pm}$ and $I_{p\Theta}$ are the chiral electric dipole - magnetic dipole and electric dipole - electric quadrupole contributions.
	Although ROA has been established first and foremost as a technique for identifying chiral molecules, the principle can be applied just as well to crystals. Nevertheless, only few results for ROA on crystalline materials have been reported, and those mostly show weak effects \cite{Intr_ROA_solid}, or acquire chirality only through external influences like magnetic field \cite{Intr_ROA_magnet} or chiral charge-density waves \cite{Intr_ROA_CDW}. More recently, a ROA-like signature in Raman spectra of NiCo$_2$TeO$_6$ was attributed to ferroaxial phonons \cite{Ferroaxial}.\\
	Ni$_3$TeO$_6$ crystallizes in the chiral and pyroelectric space group $R3$ \cite{CrystalStructure1,CrystalStructure2,Cheong1,MagneticOrder}, with planes of edge-sharing NiO$_6$ and TeO$_6$ octahedra that are stacked in a spiral pattern along the c axis. Below $T_N=52$~K \cite{MagneticOrder,MagneticOrder2}, collinear antiferromagnetic order emerges and is accompanied by a spin-induced electric polarization change \cite{MagneticOrder2,Kim2015NTO}; related first-principles work on the spin-flop transition in the isostructural Ni$_2$InSbO$_6$ is reported in Ref.~\cite{Ono2024Kinks}. Here, we show that in this material, some phonons display a ROA that is orders of magnitude higher than what has been reported for naturally chiral solid state systems. Additionally, selection rules for the chiral and achiral scattering contributions allow for complete separation of the two components in the Raman spectrum for one phonon type by controlling the polarization of incident and scattered light.\\
    From the modeling perspective, although tools to compute ROA for molecules exist \cite{ROA_computation_1,ROA_computation_2,ROA_computation_3,ROA_computation_4}, the situation is fundamentally different for a periodic solid: Optical responses are intimately connected to the quantum geometry and topology of the band structure. Transition dipole moments are directly related to the components of the quantum geometric tensor, while the magnetic dipole and electric quadrupole moments are directly related to the interband Berry curvature. Berry-geometric effects are most commonly discussed in connection with
electronic transport, where avoided crossings close to the Fermi level
often dominate intrinsic responses such as the anomalous Hall effect.
In an insulating optical response, however, there is no Fermi surface:
the relevant geometric information is carried by virtual transitions
between occupied and empty bands over a finite energy window. The
present results demonstrate that phonons can strongly modulate these
interband geometric matrix elements and the accompanying
band-dispersion terms, making Raman optical activity a sensitive probe
of electronic band geometry in an insulator. 
The splitting of chiral $E$ phonons along the chiral $c$ axis contributes to the ROA of the $E$ modes.
    
\begin{figure}[t]
		\centering
		\includegraphics[width=0.5\textwidth]{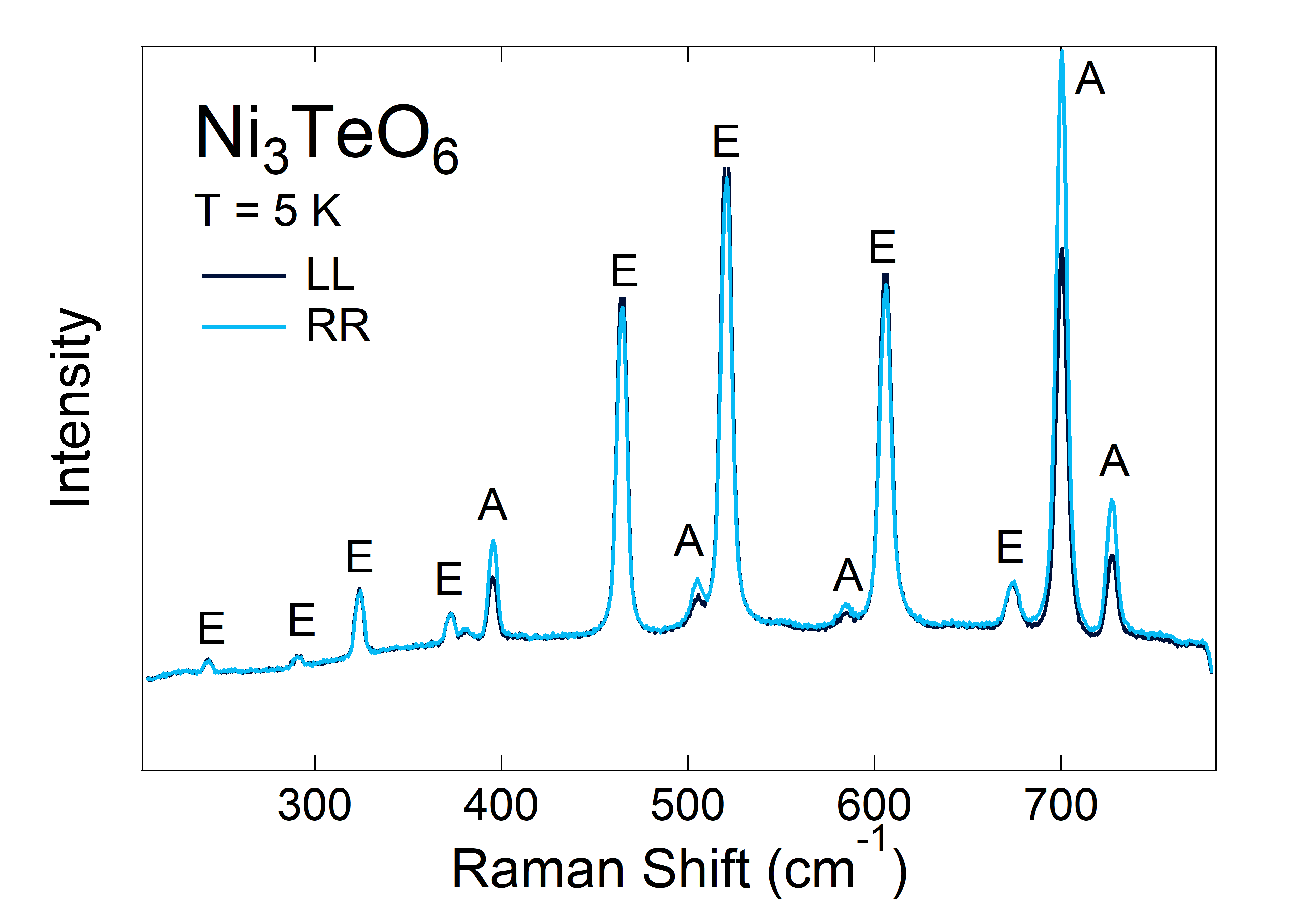}
		\caption{
        Raman spectra of $\mathrm{Ni_3TeO_6}$ at $T$ = 5 K in $(LL)$ (black) and $(RR)$ (blue) circular polarization configurations. Labels A/E refer to phonon irreducible representation. A careful peak assignment has been achieved using linear and cross-circular scattering geometries (see Supplementary Information for details).}
		\label{Spectra_5K}
	\end{figure}
	\section{Results}
	\subsection{Raman- and Raman Optical Activity Spectra}
	Selection rules for electric dipole--dipole transitions to excited phonon states are tabulated \cite{Bilbao}. For Ni$_3$TeO$_6$, they predict 27 Raman-active phonons; 9 of these are vibrations with the normal coordinate along the $c$ axis of the crystal ($A$ modes). The remaining 18 have their normal coordinate in the $ab$ plane and can be grouped into 9 pairs of degenerate phonons ($E$ modes). A thorough assignment of this symmetry to the measured Raman peaks has been achieved using measurements with linearly polarized light (see Supplementary Information and Ref.~\cite{Skiadopoulou}). In a measurement with circularly polarized light, the achiral electric dipole--electric dipole scattering $I_{pp}$ of the \textit{A} modes is allowed exclusively in cross-circular (LR and RL) geometries, and that of the \textit{E} modes exclusively in co-circular (LL and RR) geometries, where the first and second letter denote the helicity of the incident and analyzed light. Since the Raman selection rules thus forbid $I_{pp}$ for the \textit{A} modes in LL and RR spectra, any \textit{A}-mode signatures in these spectra can be interpreted as stemming exclusively from the chiral $I_{pm}$ and $I_{p\Theta}$ scattering contributions. A simple polarization-selective measurement therefore allows for a complete separation of pure electric dipole scattering and higher-order multipole scattering for the \textit{A} modes, providing an easy way to analyze the latter without an influence of the achiral scattering. This is why we focus on the co-circular LL and RR spectra below.\newline
	The difference in scattering intensity for the two circular polarizations is visualized by calculating the Raman Optical Activity
	\begin{equation}
		\centering
		\Delta  = \frac{I_R - I_L}{I_R + I_L},
		\label{ROA_eq}
	\end{equation}
	where the index R/L refers to the right- or left-circular polarization of incident light. The helicity of analyzed light can be the same as that of the incident light (co-circular ROA, $\Delta_{\rm co}$) or opposite (cross-circular ROA, $\Delta_{\rm cross}$).\\
	Theoretical considerations \cite{DCP} predict that a difference in scattering intensity for incident L/R, and thus finite ROA, can only occur for measurements in co-circular geometries ($\Delta_{\rm co}\neq0$). For cross-circular geometries, all chiral contributions to the scattering intensity should disappear, and thus the spectra measured in RL and LR configuration should be identical ($\Delta_{\rm cross}=0$).
	\begin{figure}[t]
		\centering
		\includegraphics[width=.5\textwidth]{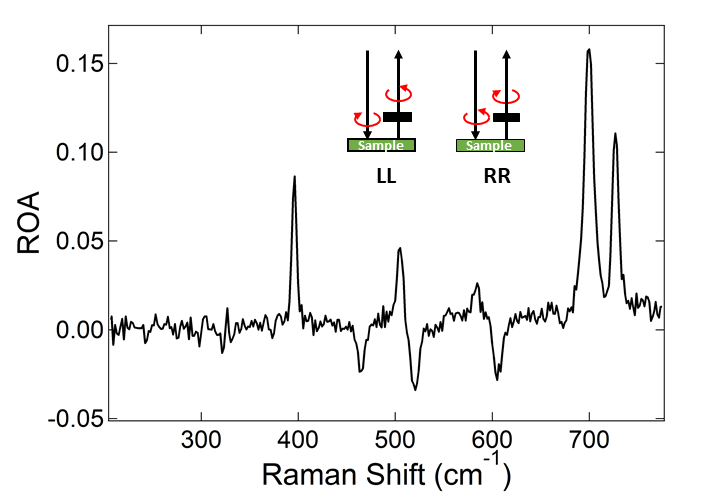}
		\caption{Co-circular ROA spectrum obtained from the spectra in Fig.~\ref{Spectra_5K} through Eq.~\eqref{ROA_eq}. Inset: Schematic depiction of co-circular LL/RR geometries.}
		\label{ROA_5K}
	\end{figure}\\
Figure~\ref{Spectra_5K} shows the Raman spectra in co-circular LL (black line) and RR (blue line) configurations. The 8 peaks at energies 243 cm$^{-1}$, 291 cm$^{-1}$, 313 cm$^{-1}$, 362 cm$^{-1}$, 464 cm$^{-1}$, 521 cm$^{-1}$, 606 cm$^{-1}$, and 674 cm$^{-1}$ are 8 of the 9 E modes expected for Ni$_3$TeO$_6$, the 9th mode likely being too weak to be observed. Surprisingly, even though the selection rules forbid achiral scattering $I_{pp}$ of the $A$ modes in co-circular geometries (see above), most of the A modes are also detectable here, albeit with significantly reduced intensity. Notably, the peak intensity of most phonons is different in RR and LL geometry. This is most pronounced for the strongest A modes at 385 cm$^{-1}$, 694 cm$^{-1}$, and 715 cm$^{-1}$. Figure~\ref{ROA_5K} shows the ROA spectrum calculated from the above spectra with Eq.~\eqref{ROA_eq}. It shows that phonons of both \textit{A} and \textit{E} symmetry contribute, because the $A$ modes always show positive ROA, while the $E$ modes show negative ROA.
    
	\begin{figure*}[hbt]
		\centering
		\includegraphics[width=\textwidth]{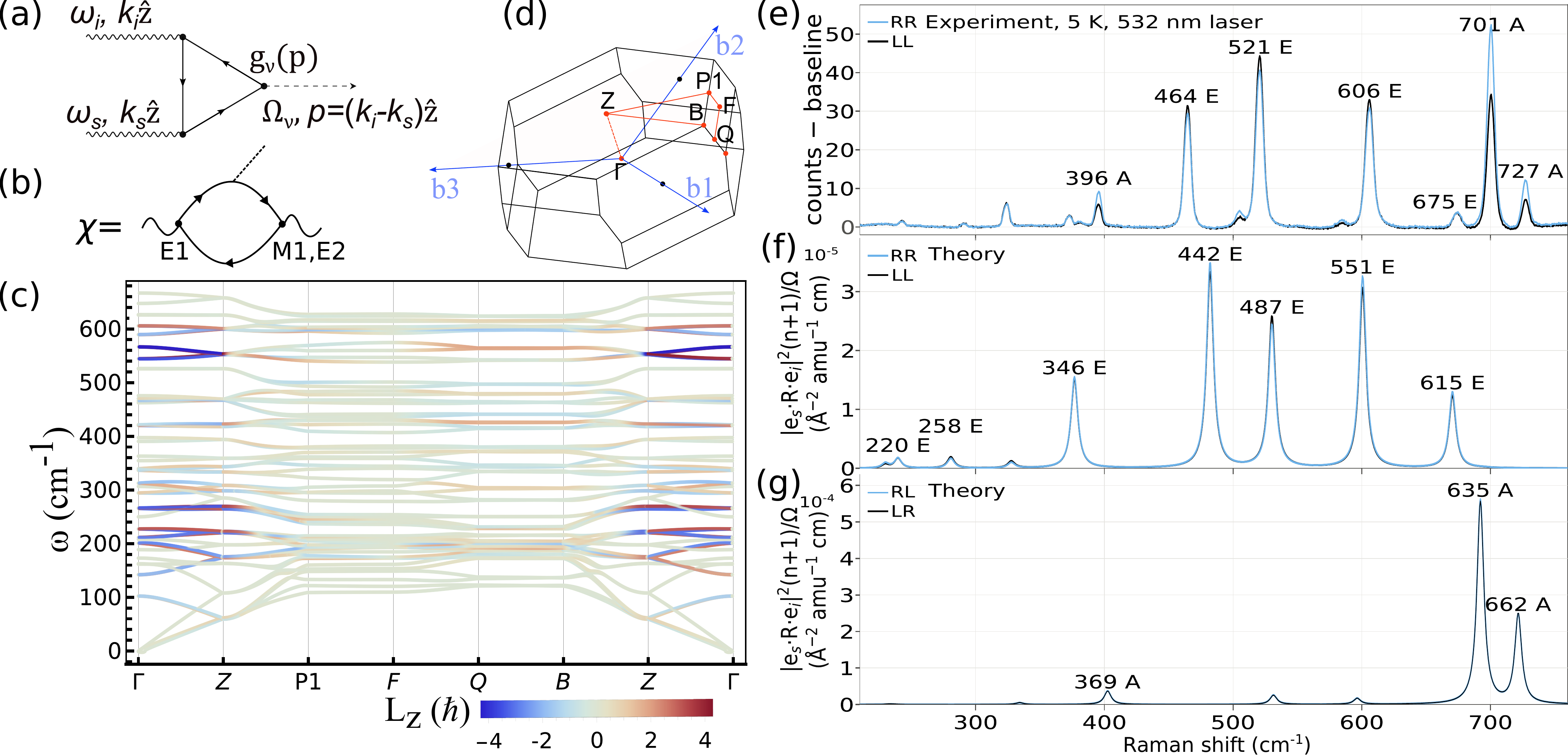}
		\caption{(a) Raman process, associated with the absorption of the incoming phonon with the momentum $\kappa_i\hat z$, emission of the phonon $p$ and a scattered photon $-\kappa_s\hat z$. Electron--phonon vertex $g_\nu(\bm p)$. (b) Contributions to Raman susceptibility $\chi$ with an electric-dipole (E1) vertex and a magnetic-dipole or electric-quadrupole (M1, E2) vertices. (c) Calculated phonon dispersion along the path through the Brillouin zone shown in (d), with the color encoding the phonon angular momentum $L_z$. (e) Measured co-circular Raman spectra at 5~K (532~nm laser, RR in blue and LL in black) with the baseline subtracted; the irreducible representations $A$ and $E$ of the phonon modes are indicated. (f) Calculated resonant Raman intensity $|\bm e_s\cdot R\cdot\bm e_i|^2(n+1)/\Omega$ of the $E$ modes in the co-circular (RR, LL) geometries. (g) The same for the $A$ modes in the cross-circular (RL, LR) geometries.}
		\label{fig:theory1}
	\end{figure*}
	The highest values are observed in the energy region between 680 cm$^{-1}$ and 720 cm$^{-1}$, peaking around 17\%. This exceeds typical vibrational ROA values of molecules, $10^{-3}$--$10^{-4}$ \cite{Barron2004book,Intr_ROA_1,Intr_ROA_2}, by two orders of magnitude, and is large also compared with natural ROA reported for enantiomorphic crystals \cite{Intr_ROA_solid}. 
    The phonons at lower energies show smaller ROA values between 1 and 5\%, albeit these are still considerably higher than values reported for any other material.\\
	Integrating the spectral peaks from the A modes in co- and cross-circular polarization measurements (see Supplementary Fig.~\ref{crosscircular}) tells us that the chiral scattering is approximately 80 times weaker than its achiral counterpart.\\
    To illustrate the temperature dependence of the effect, the development of ROA with increasing temperature is plotted in Fig.~\ref{fig:Tdep} (here plotted as the ROA of the integrated response around the strongest ROA peak at 695 cm$^{-1}$). Within the magnetically ordered phase, the ROA magnitude stays approximately the same, slightly increasing from 22\% at 5 K to 24\% at 40 K. Around the transition to the paramagnetic state at $T_N=52$~K, it shows a sharp jump up to 30\%, consistent with the strong spin--lattice coupling of Ni$_3$TeO$_6$ seen in infrared phonon spectra \cite{Yokosuk1,Yokosuk2016}. This corresponds to an increase by a factor 1.25. From $T_N$ to about 100~K, the integrated ROA is constant within error bars. For higher temperatures ($T>100$~K), determination becomes more difficult and the error of the process increases because of the declining intensity of the spectral features.

\subsection{Modelling}

\subsubsection{Kinematics of backscattering ROA}
Raman scattering involves incident $(\omega_i,\bm k_i)$ and scattered $(\omega_s,\bm k_s)$ photons and a phonon, as seen in Fig 4(a). To first order in the small photon momenta, and in the adiabatic (Placzek) limit, the Raman susceptibility of mode $\nu$ is expanded as
\begin{equation}
  \chi^\nu_{jk}(\bm k_i,\bm k_s)=R^\nu_{jk}+A^\nu_{jkl}\,k_{i,l}+B^\nu_{jkl}\,k_{s,l}.
  \label{eq:chiexp}
\end{equation}
Onsager reciprocity exchanges the incident and scattered photons and reverses their momenta,
\begin{equation}
  \chi^\nu_{jk}(\bm k_i,\bm k_s)=\chi^\nu_{kj}(-\bm k_s,-\bm k_i),
  \label{eq:onsager}
\end{equation}
where the accompanying reversal of the magnetic order parameter is compensated by the $\{T|t\}$ symmetry in the antiferromagnetic phase. Inserting Eq.~\eqref{eq:chiexp} into both sides of Eq.~\eqref{eq:onsager} and equating the terms linear in $\bm k_i$ and in $\bm k_s$ gives $R^\nu_{jk}=R^\nu_{kj}$ and
$  A^\nu_{jkl}=-B^\nu_{kjl}$. Hence $\chi^\nu_{jk}=R^\nu_{jk}+A^\nu_{jkl}k_{i,l}-A^\nu_{kjl}k_{s,l}$. Introducing the phonon momentum $\bm p=\bm k_i-\bm k_s$ (see Fig. 4(a)) and the mean photon momentum $\bm K=(\bm k_i+\bm k_s)/2$, so that $\bm k_{i,s}=\bm K\pm\bm p/2$, this becomes
\begin{equation}
  \chi^\nu_{jk}=R^\nu_{jk}+iG^\nu_{jkl}K_l+iD^\nu_{jkl}p_l,
  \label{eq:GD}
\end{equation}
where   $iG^\nu_{jkl}=A^\nu_{jkl}-A^\nu_{kjl}$,   $2iD^\nu_{jkl}=A^\nu_{jkl}+A^\nu_{kjl}$, so that $G^\nu$ is antisymmetric and $D^\nu$ symmetric in $jk$ (SI Sec.~II). The antisymmetric coefficient is the phonon derivative of the natural-optical-activity (gyrotropic) tensor, $G^\nu=\partial\gamma/\partial Q_\nu$ (related to $p=0$ version of Raman tensor, Fig. 4(b)), which is what frozen-phonon calculations of optical gyrotropy deliver \cite{Ivo10,PozoSousa23,tsirkin18gyrotropic}. It controls \emph{forward} scattering, where $\bm K\simeq\bm k$ and $\bm p\simeq0$. In \emph{backscattering} $\bm K$ is reduced to the Stokes mismatch ($\approx1\%$ of $p$ here) and $\bm p\simeq2\bm k$: the Raman analogue of optical rotation unwinding on retroreflection. The experiment therefore measures $D^\nu=-i\,\partial\chi^\nu/\partial p$, the dependence of the Raman tensor on the phonon momentum, with $p=5.9\times10^{-3}$~\AA$^{-1}$ at 2.41~eV.

 We now discuss the selection rules for backscattering ROA. A symmetric tensor contributes to the cross-circular channel only through its in-plane trace, which cannot distinguish R from L. Natural backscattering ROA therefore lives on the $E$ modes, in the co-circular channel. For $A$ modes the co-circular intensity is forbidden for light along $c$; their only intrinsic first-order CID is the Stokes-mismatch $G$ term, $\sim4\times10^{-5}$.

 The mechanisms that lead to ROA in backscattering are the following. $D^\nu$ has (i) a phonon part, from the momentum dependence of the phonon eigenvector, and (ii) an electronic part at fixed eigenvector. The phonon part contains the circular splitting of each $E$ doublet along the chiral axis, $\omega_\pm=\omega_0\pm\alpha_\nu p$ [Fig.~\ref{fig:theory1}(c)]: one handedness is created in RR, the other in LL, which gives a derivative-shaped CID with peak value $2\alpha_\nu p/{\rm FWHM}$. It also contains the admixture of other branches. The electronic part contains the first moment of the electron--phonon vertex about the electronic transition it drives (phonon-gradient term) and the shift of the electronic states and photon vertices within the scattering loop (state-shift term). The latter contains the crystalline counterpart of the molecular E1--M1 and E1--E2 interference. Both give peak-shaped CID.
    
 \subsubsection{First-principles calculations} We compute all terms from collinear antiferromagnetic DFT$+U$ and density-functional perturbation theory (DFPT$+U$) \cite{QE-2017,QE20,dudarev,GonzeLee1997}, with the electron--phonon vertex in a Wannier representation \cite{EPW2023,Marzari2012} that resolves the phonon momentum over three layers along $c$ (see Supplementary Information for computational details). The Raman tensor at finite phonon momentum is evaluated directly, in the independent-particle approximation (self-consistent local-field corrections, which can change the natural optical activity strongly \cite{ZabaloStengel23}, are not included), by coupling Bloch states at $\bar{\bm k}$ and $\bar{\bm k}\pm\bm p/2$ through the phonon; this contains the gradient, state-shift and photon--phonon contact terms at once. The resonant (finite-$\Omega$) response is obtained from a Keldysh density-matrix calculation with the $q=0$ vertex, which differs from the static calculation through the finite phonon frequency $\Omega$ in the energy denominators (Stokes versus anti-Stokes) and, for the polar LO modes, through the Fr\"ohlich coupling, which is included as an additional order of perturbation theory. Its $p\to0$ limit reproduces the frozen-phonon Raman tensors, which in turn reproduce the relative intensities of the strong $E$ lines within 30\% (SI Secs.~III--V). We add the phonon dispersion terms, treating near-degenerate $E$ pairs exactly, and construct the co-circular spectra.

 \begin{figure}[t]
    \centering
    \includegraphics[width=0.36\textwidth]{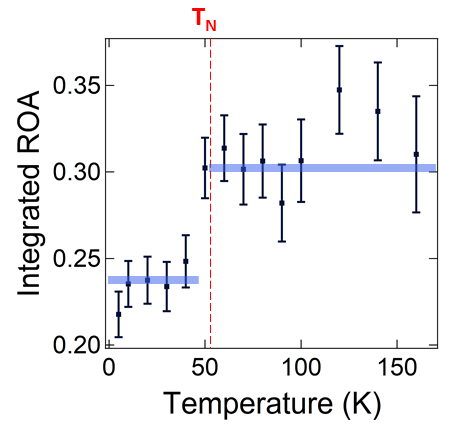}
    \caption{Temperature dependence of ROA for the most intense Raman mode around 695~cm$^{-1}$.}
    \label{fig:Tdep}
\end{figure}

The phonon dispersion for Ni$_3$TeO$_6$ is shown in Fig.~\ref{fig:theory1}(c); computational details are given in the Supplementary Information. The color of phonon branches encodes the phonon angular momentum $L_z=[\bm{e}\times \dot{\bm{e}}]_z$, where $\bm e$ is the phonon eigenvector containing mass-renormalized ionic displacements and $\dot{\bm e}$ are the corresponding velocities. Chiral phonons, representing circular rotations of ions within the $xy$ plane, clockwise and counterclockwise, are highlighted in red and blue.

The phonon and electronic parts of CID are of the same order, as seen in Fig.~\ref{fig:theory} in the Supplementary Information. The splitting gives $|\alpha_\nu|=5$--15~cm$^{-1}$\,\AA\ for the strong $E$ modes, i.e.\ $\omega_+-\omega_-=0.06$--0.17~cm$^{-1}$ at the laser energy, and CID extrema of $\pm1\%$ across the 465 and 521~cm$^{-1}$ lines. The gradient and state-shift terms partially cancel; their net peak-shaped CID ranges from $4\times10^{-4}$ (465~cm$^{-1}$) to $-1.1\times10^{-2}$ (674~cm$^{-1}$). Across the phonon lines, the total CID reaches $\pm1$--$1.5\times10^{-2}$ (SI Table~S4). This is comparable to the measured feature at 460~cm$^{-1}$ ($-1.5\%$).

We now discuss the resonant contribution to Raman intensity due to $d$--$d$ transitions \cite{Yokosuk2016}. The independent-particle approximation contains no Ni$^{2+}$ $d$--$d$ excitations, which are magnetic-dipole allowed and could enhance E1--M1 interference. Embedded-cluster CASPT2 calculations using OpenMolcas \cite{OpenMolcas2019} for all three Ni sites, differentiated along the same phonon modes (SI Sec.~X), give a $d$--$d$ contribution of $\approx10^{-4}$ to the $E$-mode CID at 2.41~eV, which lies in the transparency window between the $^3T_1(F)$ and singlet bands. On the $^3T_2$ resonance near 1.1~eV, the $d$--$d$ contribution rises to $10^{-2}$--$10^{-1}$, so excitation at 1064~nm should give a large ROA with mode-dependent signs. However, the absolute Raman intensity remains small because the $d$--$d$ transitions are very weak: the calculated Huang--Rhys factors are $\approx0.25$ and the oscillator strengths $\approx10^{-7}$. The resonant Raman calculation described above therefore captures the resonant enhancement from the band edge but not that from the $d$--$d$ excitations.

Next, we discuss the link between ROA and band geometry. Insulating optical responses have recently been discussed in terms of Berry connections and curvatures~\cite{riemannian22}, and the phonons of Ni$_3$TeO$_6$ visibly rearrange the near-degenerate valence bands at the top of the valence band. The integrand with M1 and E2 vertices \cite{PozoSousa23} involves products of three Berry connections, which for an individual band diverge as $1/\delta$ where two valence bands approach within $\delta$. These bands are, however, both occupied: the divergent terms of the two partners cancel at each $\bm k$, leaving a finite contribution controlled by the 1.7~eV gap. The Berry curvatures of individual quasi-degenerate valence bands near $\Gamma$ exceed that of the occupied manifold by orders of magnitude and change strongly with the phonon, whereas the manifold total changes by only $\sim2\%$; the integrands of $R^\nu$ and $D^\nu$ have no localized hot spots (SI Sec.~\ref{sec:bandgeom}). Therefore, the relevant band geometry is the interband (valence--conduction) connection and its modulation by the phonon~\cite{Maranzana2026PhaseSpace}. The modulation of the top valence bands by two $A$ modes is illustrated in the Supplementary Fig.~\ref{phononcell}.

We now turn to the $A$-mode signal in the co-circular spectra. The $A$ peaks and their large normalized asymmetry in the co-circular spectra are not described by the intrinsic first-order mechanisms above, which give at most $\sim10^{-4}$. Polarization leakage of $\approx2$--5\% reproduces the $A$ intensities, but it is equal in RR and LL and therefore cannot by itself produce an asymmetry; geometric tilts are negligible. We assign the remaining discrepancy between calculation and experiment to E1--E2 and E1--M1 contributions beyond those captured by the independent-particle calculation; for E1--M1, the $d$--$d$ multiplets discussed above are a natural candidate. Consistent with such an origin, the asymmetry reverses with the enantiomer and jumps at $T_N$ as seen in Fig.\ref{fig:Tdep}. A quantitative account is left for future work (SI Sec.~XI).

The calculated co-circular CID of the $E$ lines is shown in Fig.~\ref{fig:theory} of the Supplementary Information, and the measured ROA spectrum in Fig.~\ref{fig:theory1}(e).

The phonon angular momentum along $\Gamma$--$Z$ (Fig.~\ref{fig:theory1}(c)) reflects the circular splitting of the $E$ doublets, which gives the phonon part of the CID (Fig.~\ref{fig:theory} in the Supplementary Information); the electronic part is of the same order.

Below $T_N$, The AFM ordering Ni ions along the three-fold rotation axis order as $\uparrow\downarrow\downarrow\downarrow\uparrow\uparrow$. The magnetic unit cell is twice as large as the structural one along the polar $c$ axis, such that the magnetic Brillouin zone has half the extent of the structural one. The $Z$ point therefore folds onto the magnetic-zone center $\Gamma_{\mathrm M}$. We therefore label the calculated modes $A_\Gamma$, $A_Z$, $E_\Gamma$, and $E_Z$ according to their wave vector in the parent crystallographic Brillouin zone. The cell doubling folds the $A$ modes at the $Z$ point to $\Gamma$ and gives rise to $\epsilon^A_{xy}\propto\mathcal{L}$, since both transform the same way under the $\{T|t\}$ symmetry that combines time reversal and a translation by one crystallographic cell along the three-fold axis. Therefore, weak lines are expected to appear below the N\'{e}el temperature, but they are not seen in the experimental data. The combined operation $\{T|t\}$ fixes the folded tensors completely: together with reciprocity in exact backscattering, $k_s=-k_i$, it requires $R_{ab}=-R_{ba}$ for every $Z$-derived mode, at any order in the photon momentum, while in a collinear calculation without spin--orbit coupling the two spin channels cancel identically. Our calculations confirm this: with SOC, the symmetric part of the $A_Z$ tensors stays at the numerical noise floor, while the antisymmetric part is finite and $\propto\mathcal L$ by symmetry. Consequently, along $c$ only the $A_Z$ modes are visible, in the cross-circular channel, with the amplitudes $\pm i\epsilon^A_{xy}$ for the two helicities, i.e.\ with equal intensities and no ROA. The $E_Z$ modes carry no antisymmetric in-plane tensor and are not observable in backscattering along $c$. Folded modes therefore cannot contribute to the co-circular ROA. The detailed symmetry argument is given in SI Sec.~\ref{sec:folded}.

\section{Discussion}
We employed circularly polarized Raman spectroscopy in backscattering geometry to measure the Raman optical activity of Ni$_3$TeO$_6$ above and below the magnetic ordering temperature $T_N=52$~K. Both $A$ and $E$-symmetric phonon modes show large intensity differences for incident left- and right-circularly polarized light, leading to peak ROA values of $\approx $ 15\%. The origin of ROA is strong, chiral magnetic dipole- and electric quadrupole transition moments. Calculations of the Raman tensor at finite phonon momentum point towards phonon-induced modification of the band geometry as its origin. As an outlook, the E1--E2 and E1--M1 contributions, to which we assign the remaining discrepancy with experiment, can be addressed by adding the off-diagonal Berry connection and intra-Wannier multipoles to the present model, and, including spin--orbit coupling. The previously unrecognized connection between band geometry and chiral scattering contributions establishes ROA, which has primarily been used as a technique to research the chirality of molecules, as a suitable probe for crystalline samples.\\
The detection of a ROA signal in a material with a magnetic phase transition also made it possible for the first time to analyze the influence of such a transition on the chiral scattering contributions. We showed that the ROA magnitude of the highest-contributing mode changes significantly at T$_N$.\\
Calculations of phonon angular momentum additionally enabled the identification of chiral phonons in Ni$_3$TeO$_6$, whose circular splitting contributes to the ROA of the $E$ modes. Backscattering ROA can now be used to probe the phonon modulation of the quantum geometry in other materials. Narrow-gap chiral or polar compounds with strong electron--phonon coupling and sizeable spin--orbit coupling are the most promising, since there the phonons and spin--orbit coupling can modify the quantum geometry much more strongly than in the wide-gap insulator Ni$_3$TeO$_6$. 
\section{Acknowledgements}
The authors would like to thank S.-W. Cheong, A. Fujimori, D.-J. Huang, and D. I. Khomskii for inspiring and useful discussions.
This work was partially funded by the Deutsche Forschungsgemeinschaft (DFG, German Research Foundation) through CRC 1238 (Project No. 277146847) and TRR360 (Project No. 492547816). S.V.S. and E.V.K. thank the Ministry of Science and Higher Education of the Russian Federation for support through funding the Institute of Metal Physics.

\section*{Author contributions}
PHMvL has conceived the project, led the measurements; PS performed the experiments and wrote the initial draft under the supervision of PHMvL; TCK, VT, IK and MG performed the experiments; RK and SA computed phonon chirality and calculated Raman intensity; EK and SVS computed phonons; all authors revised the manuscript.
\section*{Competing interests}
The authors declare no competing financial or non-financial interests.

\section{Methods}
	\subsection{Samples}
	
	High-quality single crystal samples of Ni$_3$TeO$_6$ were grown as described in \cite{David} using the chemical transport reaction method and subsequently characterized by x-ray diffraction and polarization microscopy. The crystals are green, transparent, hexagonal plates with a diameter of $\approx $ 1 mm.
	\subsection{Raman measurements}
	The sample was mounted in a liquid helium flow cryostat. Measurements were performed on a polished (\mgdel{ab}\mg{$ab$}) surface in backscattering geometry using a 532 nm solid state laser with a power of 1~mW at the sample and a spot size of 30 $\mathrm{\mu m}$. The wavelength lies within a small energy window where the absorption of the sample is minimal \cite{DirectionalDichroism}. The linearly polarized light from the laser is converted to circular polarization by a quarter-wave plate in front of the sample, and the backscattered light is converted back to linearly polarized light by the same optics. A linear polarizer is then used to select the polarization to be analyzed. This allows for measurement of all 4 dual circular polarization configurations LL, RR, LR\mg{,} and RL, where the first and second letter denote the helicity of the incident and analyzed light, respectively. This setup corresponds to the backscattering dual circular polarization geometry that has been shown to detect ROA most efficiently and with the least experimental artifacts \cite{backscatteringROA}.
	Spectra were recorded using a Jobin Yvon T64000 triple grating Raman spectrometer equipped with a liquid nitrogen cooled CCD yielding a total energy resolution well below 1 meV. Since there are no phonons at energies below 170 cm$^{-1}$, we restrict ourselves to the energy range between 160 cm$^{-1}$ \  and 750 cm$^{-1}$, thereby covering the entire phonon energy range of this material. Measurements in RR and LL geometry were done at various temperatures between 5 K and 200 K, with 10 K steps between 10 and 100 K and 20 K steps above. The integration time per spectrum was 10 minutes. Additionally, measurements with exposure times of 30 minutes were done in LR and RL geometry at room temperature.
\bibliography{references_V22}
\clearpage
\widetext
\section{Supplementary Information}
\renewcommand{\thefigure}{S\arabic{figure}}
\setcounter{figure}{0}

\subsection{Raman spectra in linear polarization geometry}
\begin{figure}[h]
    \includegraphics[width=0.5\textwidth]{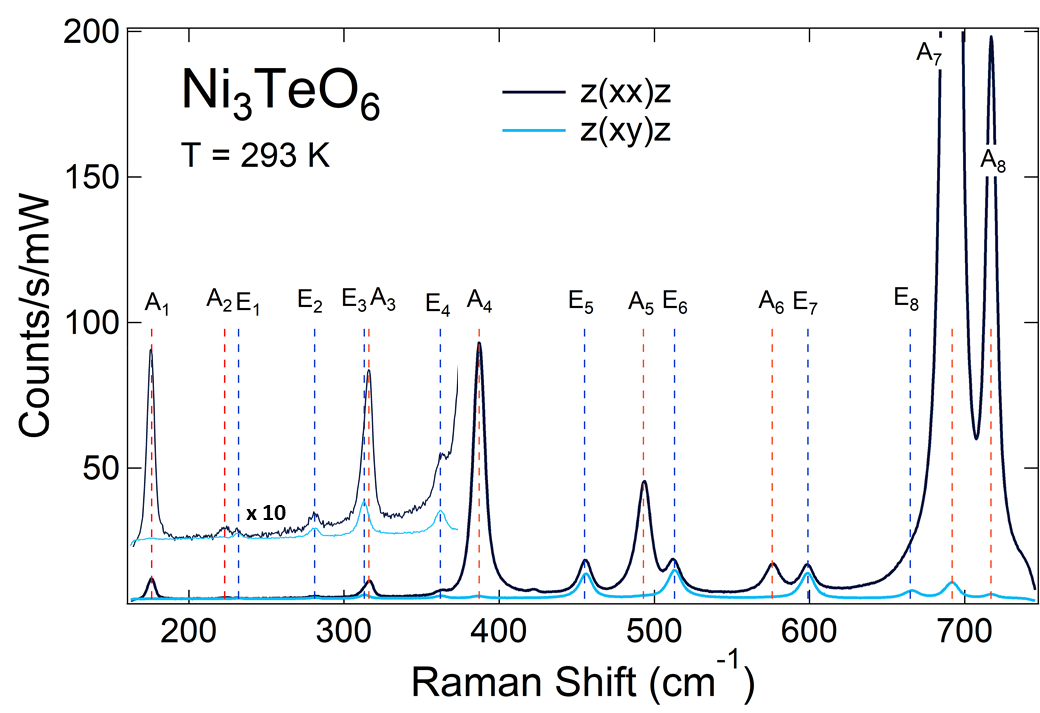}
    \caption{Raman spectra of Ni$_3$TeO$_6$ measured in linear $z(xx)\bar{z}$ and $z(xy)\bar{z}$ geometries. Labels indicate phonon representation.}
    \label{LinearPol}
\end{figure}

In order to be able to differentiate between \textit{A} and \textit{E} modes, a Raman measurement with linearly polarized incoming and analyzed light has been carried out. The irreducible representation for R3 space group is $\Gamma_{Raman}\ = 9\ A\ + 9\ E(x)\ +\ 9\ E(y)$. According to the selection rules, the 9 \textit{A} modes should only be detectable in $z(xx)\bar{z}$ geometry, while the 9 degenerate pairs of E modes should be active in both $z(xx)\bar{z}$ and $z(xy)\bar{z}$ geometries. The result of these measurements at room temperature is shown in Fig.~\ref{LinearPol}. The incident laser power was 1 mW, and the measurement time was 30 minutes for $z(xx)\bar{z}$ and 75 minutes for $z(xy)\bar{z}$ geometries. The broadness of the peak in the co-circular ROA spectrum around 690~cm$^{-1}$ suggests the presence of an additional \textit{A} mode at 691~cm$^{-1}$, which would complete the set of \textit{A} modes expected for this space group.
\subsection{Raman spectra in cross-circular geometries}
\begin{figure}[h]
    \includegraphics[width=0.5\textwidth]{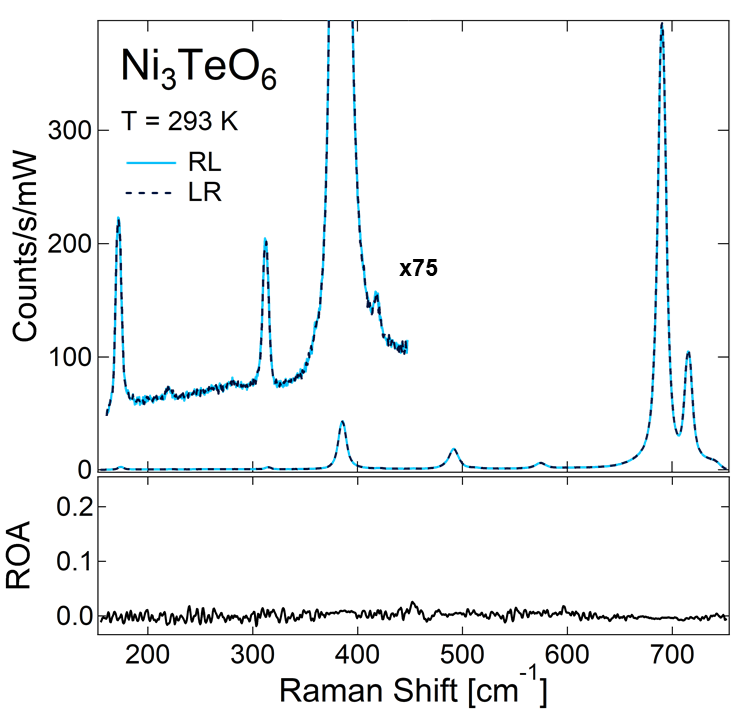}
    \caption{Top: Raman spectra of Ni$_3$TeO$_6$ in RL and LR configurations. The \textit{A} modes known from Fig.~\ref{LinearPol} are visible, and there is no intensity difference for LR/RL. Bottom: The ROA spectrum obtained for this geometry is flat.}
    \label{crosscircular}
\end{figure}

In cross-circular geometries, i.e., RL and LR configuration, all 9 \textit{A} modes should be active, while the 9 pairs of degenerate \textit{E} modes are forbidden. Additionally, there should be no cross-circular Raman optical activity. The Raman spectra in RL and LR configuration are shown in Fig.~\ref{crosscircular}. As expected, they show the 8 $A$ modes known from Fig.~\ref{LinearPol}, with no intensity differences between LR and RL, and accordingly no cross-circular ROA.

\subsection{Sign change in ROA for different domains}
\begin{figure}[h]
    \includegraphics[width=0.5\textwidth]{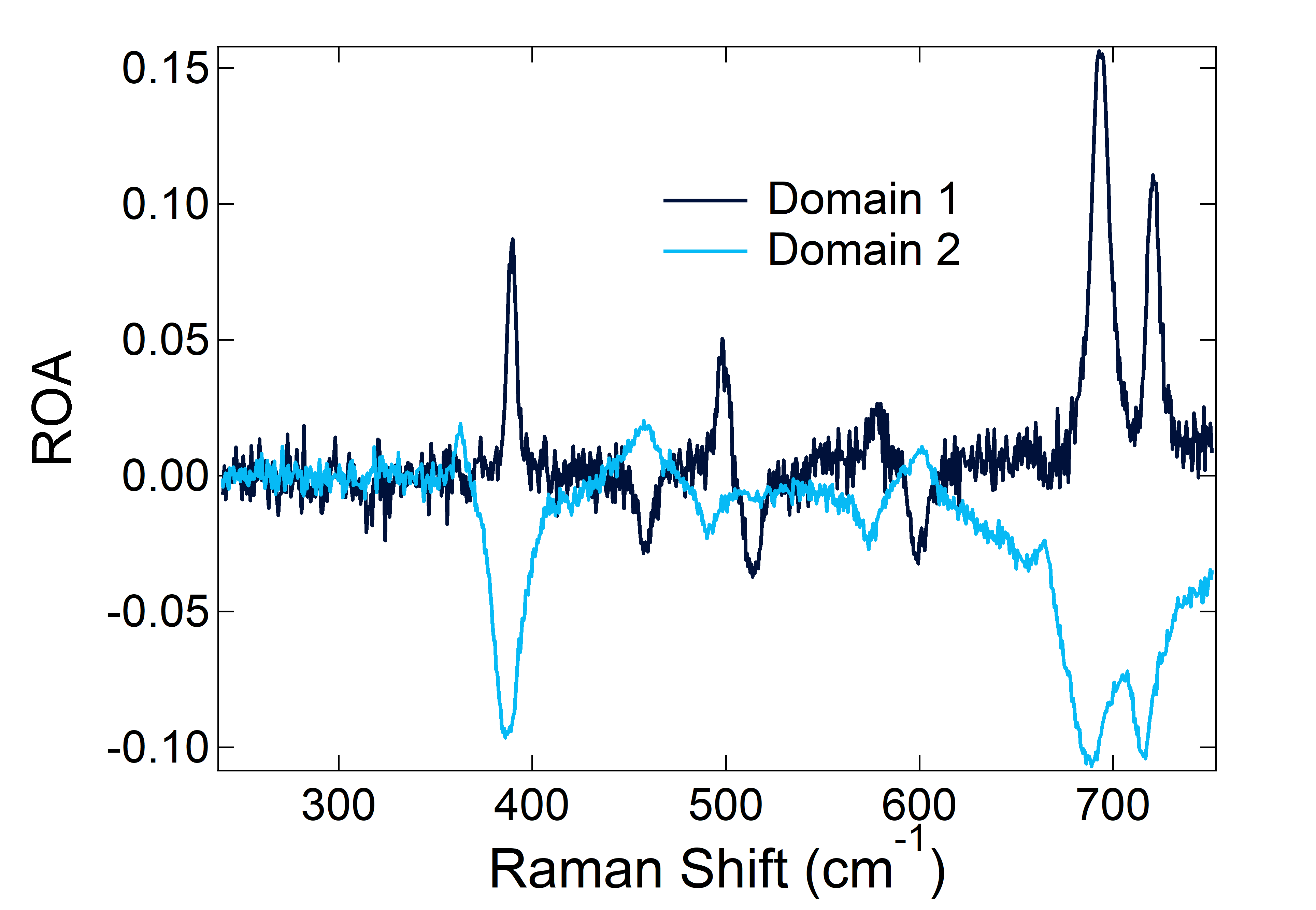}
    \caption{Co-circular ROA spectra of Ni$_3$TeO$_6$ measured in the L- and R chiral domains. The spectra show the same features, but with opposite sign.}
    \label{ROA_domains}
\end{figure}

The sample of Ni$_3$TeO$_6$ used for the measurements in this work shows well separated domains of the L- and R-variant of the crystal structure. While all measurements shown in the main part of this article were measured on the same domain, measurements on the other domain were also done and are shown in Fig.~\ref{ROA_domains}. As expected, the sign of ROA is opposite for the other domain type in the sample.

\subsection{Calculated co-circular CID of the $E$ lines}
\begin{figure}[h]
    \centering
    \includegraphics[width=0.7\textwidth]{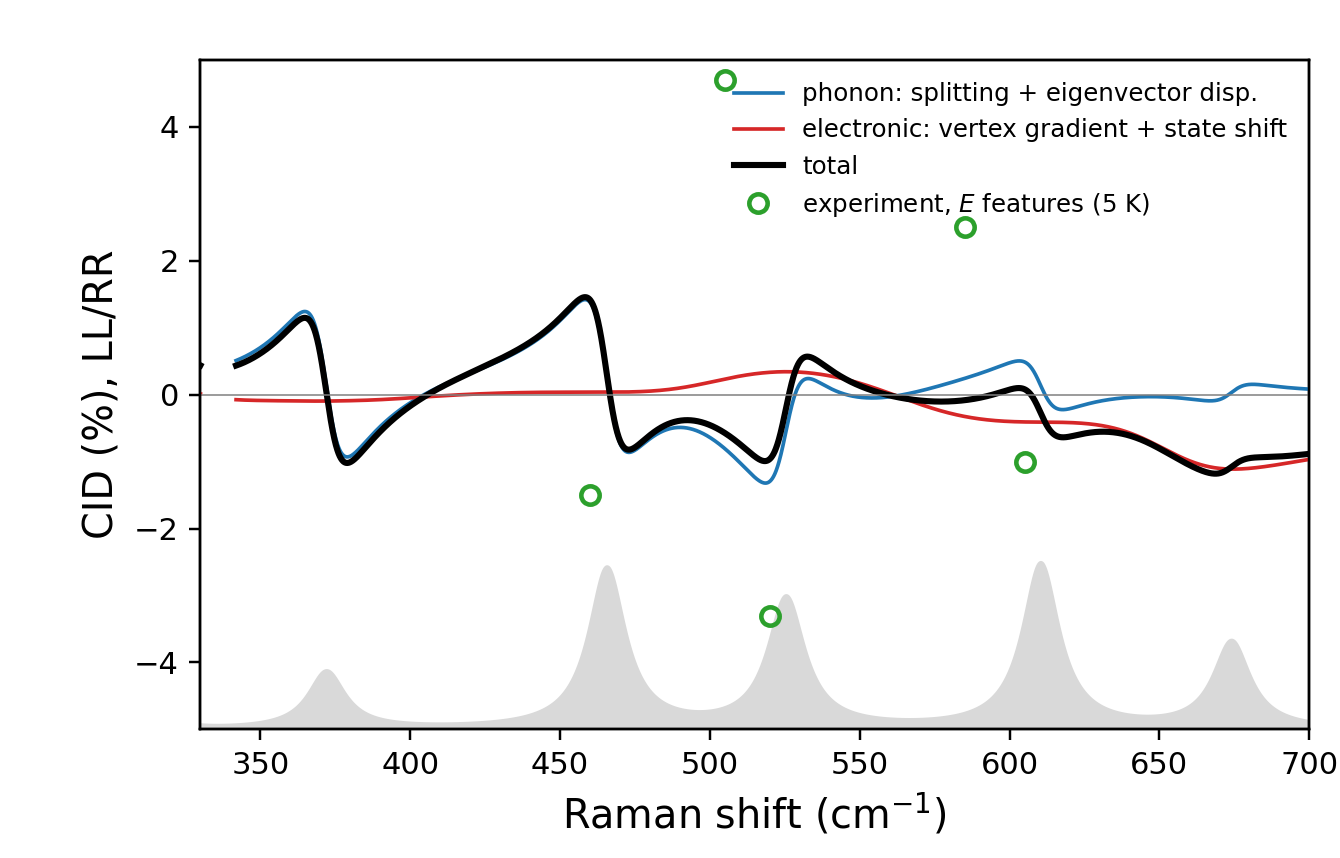}
    \caption{Calculated co-circular CID of the $E$ lines (all first-order terms, FWHM 15~cm$^{-1}$, calculated lines placed at the assigned experimental frequencies), its phonon (splitting + eigenvector dispersion) and electronic (vertex gradient + state shift) parts, and the features of the measured ROA spectrum. The circles mark the extrema of the measured derivative-shaped features around the $E$ lines (e.g.\ 505 and 520~cm$^{-1}$ around the 521~cm$^{-1}$ line). The $A$ lines (e.g.\ at 396~cm$^{-1}$ and above 700~cm$^{-1}$) are not part of this calculation: their intrinsic co-circular CID is negligible, and their intensity in this channel is modelled as leakage, which gives no CID. The overall sign of the calculation is not fixed.}
    \label{fig:theory}
\end{figure}

\subsection{Folded modes below $T_N$ in backscattering along $c$}\label{sec:folded}
Below $T_N$ the magnetic cell is doubled along $c$, and the phonons at the crystallographic $Z$ point fold onto the magnetic zone center. The collinear order preserves the symmetry $\{T|t\}$, time reversal combined with the translation $t=\bm c_{\rm hex}$, which reverses $\mathcal L$ twice. Onsager reciprocity by itself relates the state with $\mathcal L$ to the state with $-\mathcal L$:
\begin{equation}
  \chi^\nu_{jk}(\bm k_i,\bm k_s;\mathcal L,Q)=\chi^\nu_{kj}(-\bm k_s,-\bm k_i;-\mathcal L,Q),
\end{equation}
where the phonon coordinate $Q$ is even under $T$. The translation $t$ maps the state with $-\mathcal L$ back to the state with $\mathcal L$ and acts on the phonon coordinate: $Q\to Q$ for a mode derived from $\Gamma$, and $Q\to-Q$ for a mode derived from $Z$, so that the two do not mix. For a $\Gamma$-derived mode this gives Eq.~\eqref{eq:onsager}. For a $Z$-derived mode the susceptibility, which is linear in $Q$ in the adiabatic (Placzek) limit, changes sign:
\begin{equation}
  \chi^\nu_{jk}(\bm k_i,\bm k_s)=-\chi^\nu_{kj}(-\bm k_s,-\bm k_i)\qquad(\nu\in Z).
\end{equation}
Inserting the expansion, Eq.~\eqref{eq:chiexp}, and equating the terms of equal order in the photon momenta gives $R^\nu_{jk}=-R^\nu_{kj}$ and $A^\nu_{jkl}=+B^\nu_{kjl}$, so that $\chi^\nu_{jk}=R^\nu_{jk}+A^\nu_{jkl}k_{i,l}+A^\nu_{kjl}k_{s,l}$. In exact backscattering ($\bm k_s=-\bm k_i=-\bm k$) this reduces to $\chi^\nu_{jk}=R^\nu_{jk}+(A^\nu_{jkl}-A^\nu_{kjl})k_l$. The tensor of every $Z$-derived mode is therefore antisymmetric in $jk$ in backscattering, at any order in the photon momentum.

\emph{$A_Z$ modes.} The product $Q(A_Z)\,\mathcal L$ of the phonon coordinate and the antiferromagnetic order parameter is odd under $T$, even under $t$ and invariant under $C_3$. It transforms like a magnetic field along $z$ and generates the antisymmetric in-plane tensor $\epsilon^A_{xy}\propto\mathcal L Q$, which is allowed by symmetry. For circular polarizations it has amplitudes $\pm i\epsilon^A_{xy}$ in the cross-circular channels RL and LR, which therefore have equal intensities, so these modes are visible but carry no ROA. In our calculation the symmetric part of the $A_Z$ tensors stays at the numerical noise floor ($\lesssim10^{-3}$ of the strongest $A_\Gamma$ tensor), while the antisymmetric part is finite.

\emph{$E_Z$ modes.} The product $Q(E_Z)\,\mathcal L$ has $E$ symmetry. A tensor antisymmetric in the plane is of $A$ type and cannot be linear in an $E$ coordinate, and the symmetric in-plane part, which would be needed in the co-circular channel, is excluded by the relation above. The $E_Z$ modes are therefore not observable in backscattering along $c$, and they cannot account for the change of the ROA at $T_N$. They could appear only through components with a $z$ index, which are not probed in this geometry (a focused beam gives these only as polarization leakage). We have not calculated the $E_Z$ tensors explicitly; this conclusion rests on symmetry. Weak $Z$-derived lines below $T_N$ are not observed.

\subsection{Band geometry and Brillouin-zone convergence}\label{sec:bandgeom}
\emph{Pairwise cancellation.} Inserting a resolution of the identity into the M1 and E2 vertices turns the E1$\times$(M1, E2) integrand into products of three Berry connections. If an occupied band $n$ has a quasi-degenerate occupied partner $n'$ at splitting $\delta$, the intermediate state $m=n'$ contributes $A_{nn'}\simeq v_{nn'}/\delta$, so the integrand of a single band diverges as $1/\delta$. The partner contributes the same term with opposite sign, because hermiticity gives $T_{n'n}=-T^*_{nn'}$ for the triple product $T_{nn'}$ of connections. For a frequency factor $F$ of the valence--conduction energy $x=E_l-E_n$, the pair sum is
\begin{equation}
  {\rm Re}\,T_{nn'}\big[F(x)(x+\delta)-F(x+\delta)\,x\big]=\delta\,{\rm Re}\,T_{nn'}\big[F(x)-xF'(x)\big]+O(\delta^2),
\end{equation}
which is finite: $\delta\,T_{nn'}$ contains only the intra-valence velocity $v_{nn'}$ and connections across the gap, and $F-xF'$ is smooth on the scale of the gap (1.74~eV). The cancellation occurs at each $\bm k$ between the two partners. Genuine hot spots of the band-summed integrand would require small valence--conduction gaps, which do not occur here.

\begin{figure}[h]
  \centering
  \includegraphics[width=\textwidth]{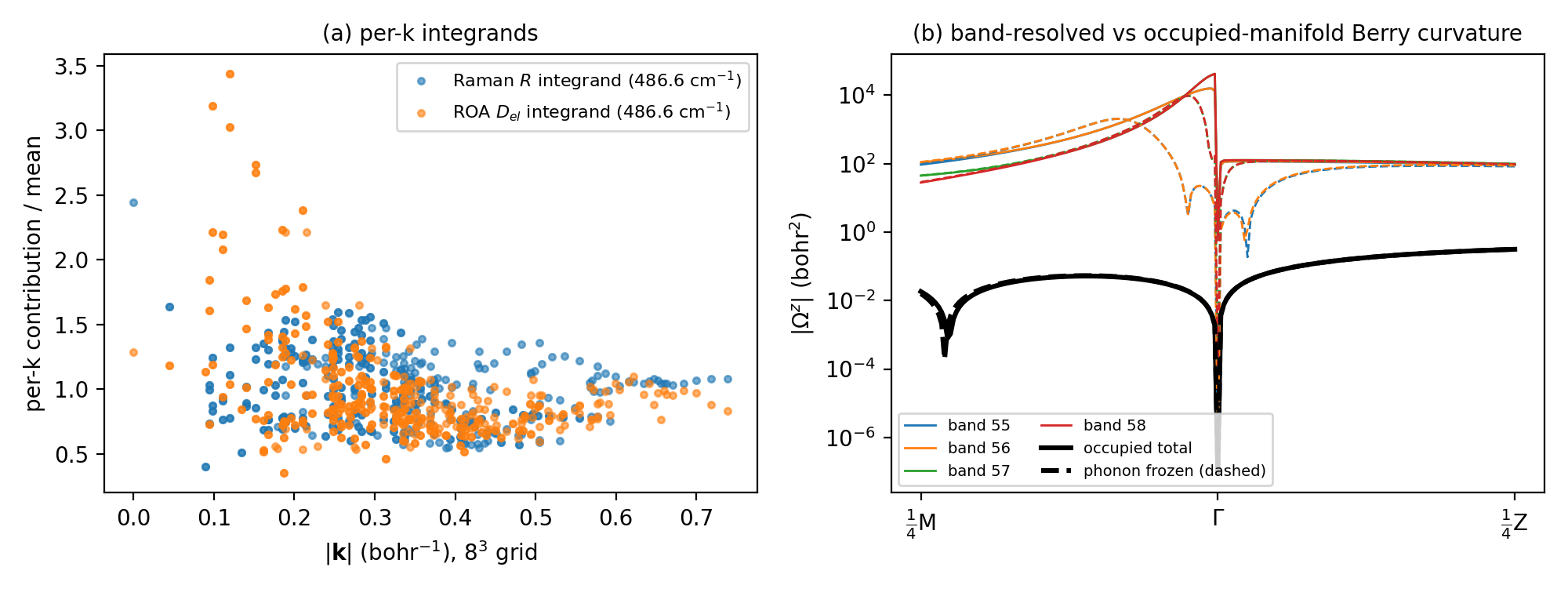}
  \caption{(a) Per-$\bm k$ contributions to the Raman tensor and to the electronic ROA coefficient of the 486.6~cm$^{-1}$ $E$ mode on the $8^3$ mesh (both spins), normalized to their mean, versus distance from $\Gamma$. (b) Berry curvature $|\Omega^z|$ of the four top valence bands (thin) and of the occupied manifold (thick) along $\frac14$M--$\Gamma$--$\frac14$Z, at equilibrium (solid) and with the 442.5~cm$^{-1}$ $E$ phonon frozen in (dashed; spin up).}
  \label{fig:kberry}
\end{figure}

\emph{Berry curvature.} Near $\Gamma$ the top valence bands form doublets split by $<1$~meV at $|\bm k|=0.002$~bohr$^{-1}$. Their individual Berry curvatures are $\Omega^z_n=\pm1.4\times10^4$ and $\pm2.9\times10^4$~bohr$^2$, while the occupied-manifold total is $5.9\times10^{-3}$~bohr$^2$ (Fig.~\ref{fig:kberry}b). Freezing the 442.5~cm$^{-1}$ $E$ phonon ($Q=0.02$~\AA$\sqrt{\rm amu}$) splits the $\Gamma$ doublets by up to 4.3~meV and changes the individual curvatures by three orders of magnitude (to $\pm9$ and $\pm170$~bohr$^2$), but changes the occupied total by only 2\% ($5.94\to6.07\times10^{-3}$~bohr$^2$). The same holds at $|\bm k|=0.01$ and 0.03~bohr$^{-1}$. At $\Gamma$ the 638.4~cm$^{-1}$ $A$ mode shifts the valence doublets rigidly ($\pm2.5$~meV) and the conduction-band edge by $\mp12.6$~meV; this deformation-potential shift of the gap, not a valence splitting, modulates the optical response.

\emph{Brillouin-zone distribution of the backscattering integrands.} The per-$\bm k$ integrands of $R^\nu$ and of the electronic part of $D^\nu$ on the $8^3$ mesh are spread over the whole zone (Fig.~\ref{fig:kberry}a). Their participation ratios are 0.67--0.94 (Raman) and 0.65--0.87 (ROA) of the number of $\bm k$ points, where 1 means uniform. Spheres of radius 0.15 (in units of the reciprocal lattice vectors) around $\Gamma$, $Z$ and the zone-boundary points $X_1$ and $X_2$ of the $c$-aligned cell carry 1--5\% of $\sum|R(\bm k)|$ and $\sum|D(\bm k)|$, close to their volume fraction (1.4\%). The ROA integrand involves strong cancellation, $|\sum_{\bm k}D(\bm k)|/\sum_{\bm k}|D(\bm k)|=0.1$--0.6, but no localized hot spot. The per-$\bm k$ distribution of $D$ is not unique, because $\bm k$-total derivatives can be moved between $\bm k$ points; only refinement tests are unambiguous.

\emph{Local refinement.} Replacing the contribution of the coarse cell around $\Gamma$, $X_1$, $Z$ and $X_2$ by an average over a $3^3$ or $5^3$ sub-grid changes the zone totals by at most $2\times10^{-3}$ (Raman) and $9\times10^{-3}$ (ROA) at $\Gamma$, $\le1.5\times10^{-4}$ and $\le9\times10^{-4}$ at $X_1$, $\le4\times10^{-4}$ and $\le6\times10^{-3}$ at $Z$, and $\le9\times10^{-5}$ and $\le5\times10^{-4}$ at $X_2$, for all strong $E$ modes and the 638.4~cm$^{-1}$ $A$ mode. The $3^3$ and $5^3$ results coincide, so there is no unresolved singularity. The backscattering ROA is therefore not controlled by the Berry geometry of individual quasi-degenerate valence bands, and needs no special sampling near $\Gamma$, $Z$ or the zone boundary.

\begin{table}[h!]
\centering
\renewcommand{\arraystretch}{1.2} 
\begin{tabular}{c c c c | c c c c}
\toprule
\multicolumn{4}{c|}{\textbf{Magnetic Unit Cell}} & \multicolumn{4}{c}{\textbf{Non-Magnetic Unit Cell}} \\
\textbf{Mode(s)} & \textbf{THz} & \textbf{cm$^{-1}$} & \textbf{Symmetry} & 
\textbf{Mode(s)} & \textbf{THz} & \textbf{cm$^{-1}$} & \textbf{Symmetry} \\
\midrule
56,57      & 3.058      & 101.9      & E$_z$      & 53         & 4.848      & 161.7      & A          \\
54,55      & 4.260      & 142.0      & E$_z$      & 49         & 6.016      & 200.5      & A          \\
52         & 4.872      & 162.5      & A$_z$      & 47,48      & 6.022      & 200.7      & E          \\
51         & 5.190      & 173.0      & A$_z$      & 45,46      & 6.340      & 211.4      & E          \\
50         & 5.660      & 188.7      & A$_z$      & 41         & 7.819      & 260.7      & A          \\
43,44      & 6.814      & 227.0      & E$_z$      & 39,40      & 7.959      & 265.5      & E          \\
42         & 7.495      & 249.9      & A$_z$      & 35,36      & 8.821      & 294.0      & E          \\
37,38      & 7.985      & 266.1      & E$_z$      & 31,32      & 10.087     & 336.5      & E          \\
33,34      & 9.371      & 312.3      & E$_z$      & 28         & 10.504     & 350.1      & A          \\
29,30      & 10.169     & 339.0      & E$_z$      & 25         & 11.914     & 397.5      & A          \\
27         & 10.622     & 354.3      & A$_z$      & 23,24      & 12.592     & 419.8      & E          \\
26         & 11.705     & 390.0      & A$_z$      & 20         & 12.631     & 421.1      & A          \\
21,22      & 12.627     & 421.0      & E$_z$      & 16,17      & 14.096     & 469.8      & E          \\
18,19      & 13.967     & 465.5      & E$_z$      & 14         & 14.969     & 498.9      & A          \\
15         & 14.256     & 475.2      & A$_z$      & 11,12      & 16.319     & 544.3      & E          \\
13         & 15.775     & 525.9      & A$_z$      & 8          & 17.522     & 584.0      & A          \\
9,10       & 16.984     & 566.0      & E$_z$      & 4,5        & 18.156     & 605.6      & E          \\
6,7        & 17.667     & 588.6      & E$_z$      & 2          & 18.992     & 633.2      & A          \\
3          & 18.782     & 626.0      & A$_z$      &            &            &            &            \\
1          & 19.987     & 666.5      & A$_z$      &            &            &            &            \\
\bottomrule
\end{tabular}
\caption{Vibrational modes of Ni$_3$TeO$_6$, separated into magnetic unit cell modes and non-magnetic unit cell modes. The assignment follows from the parity $\langle e|P_t|e\rangle=\pm1$ of each eigenvector under the structural translation $t=\mathbf c_{\rm hex}$ (which exchanges the two antiparallel Ni sublattices and is lost in the doubled cell): $+1$ for $\Gamma$-point modes of the crystallographic cell, $-1$ for modes folded from $Z$ (subscript $z$); all parities are $\pm1.000$ except the 161.7/162.5 pair ($\pm0.97$). This gives 9$A$+9$E$ optical $\Gamma$ modes, as required for $R3$, and 10$A_z$+10$E_z$. Four entries listed before as crystallographic $\Gamma$ modes (390.0, 354.3, 249.9, 101.9~cm$^{-1}$) are folded $Z$ modes.}
\label{tab:combined_modes_with_cm}
\end{table}

\subsection{Computational details}
Electronic-structure, phonon and electron--phonon calculations were performed with Quantum ESPRESSO \cite{QE-2009,QE-2017,QE20} and EPW \cite{EPW2023}; only the phonon dispersion and phonon angular momentum shown in Fig.~\ref{fig:theory1}(c) were obtained with DFPT \cite{GonzeLee1997} in the VASP code \cite{KresseHafner1993,KresseHafner1994,KresseFurthmuller1996CMS,KresseFurthmuller1996PRB,KresseHafner1994USPP,KresseJoubert1999}.

\emph{Electronic structure.} We used collinear spin-polarized LSDA \cite{Ceperley,Perdew} with optimal norm-conserving Vanderbilt (ONCV) pseudopotentials in the SG15 version of Schlipf and Gygi \cite{Hamann13,Schlipf15}, plane-wave cutoffs of 60~Ry (wave functions) and 240~Ry (charge density), no spin--orbit coupling, and the simplified LSDA+$U$ method \cite{dudarev} with $U=6.5$~eV on the Ni-$3d$ states (atomic projectors). The self-consistent runs used a shifted $4\times4\times4$ $k$-grid. The antiferromagnetic cell contains 6 Ni, 2 Te and 12 O ions (20 atoms); it was re-based such that its third reciprocal lattice vector is parallel to $c$. The structure was relaxed until the forces were below $10^{-4}$~Ry/Bohr.

\begin{figure*}[hbt]
	\centering
	\includegraphics[width=\textwidth]{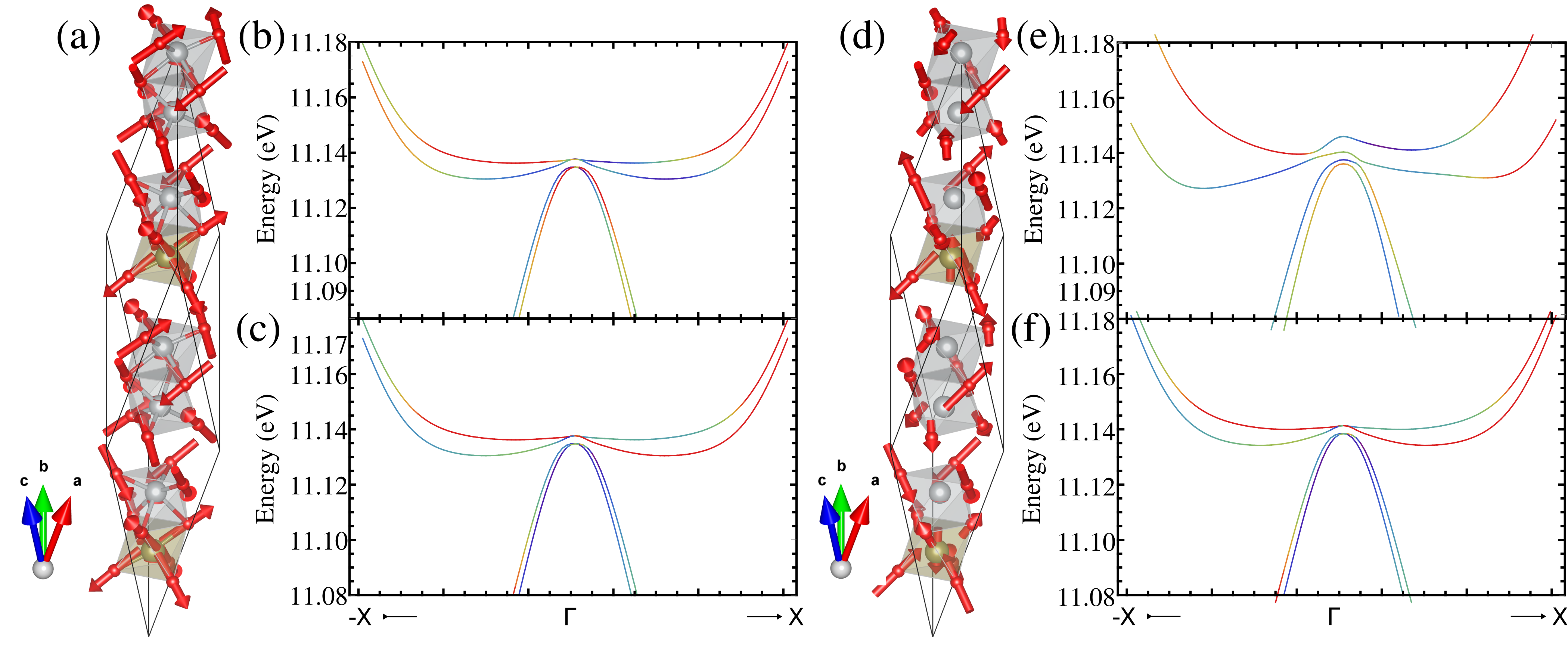}
	\caption{Phonon-induced modification of the top valence bands near $\Gamma$. (a) Ionic displacements (red arrows) of the second-highest ($A_\Gamma$) mode and (d) of the highest ($A_Z$) mode in the magnetic cell. (b) Electronic bands along $-X$--$\Gamma$--$X$ with second-highest frequency mode ($A_\Gamma$ mode) frozen in and (c) in the equilibrium structure with the colors encoding the first principal component of the character change with $k$ in the top band; (d) the highest frequency ($A_z$) mode and (e) the corresponding  bands with the mode frozen in, with colors encoding the principal component of the character change with $k$ in the top band, and (f) the equilibrium bands with the same color coding. The phonons change the separations and the hybridization of the near-degenerate top valence bands near $\Gamma$.}
	\label{phononcell}
\end{figure*}

\emph{Phonons.} Phonons were computed in the 20-atom magnetic cell with DFPT+$U$ on a $1\times1\times3$ $q$-grid along $c$ ($\Gamma$ and $\pm\frac{1}{3}$ of the $c$-axis reciprocal vector); the dielectric tensor and Born effective charges were computed at $\Gamma$ only. The transverse-optical (TO) eigenvectors at $\Gamma$ were used and matched to the VASP set by their overlap.

\emph{Electron--phonon vertex.} EPW \cite{EPW2023} was run separately for each spin on a coarse $4\times4\times6$ $k$-grid and the $1\times1\times3$ $q$-grid. We used 66 Wannier functions per spin \cite{Marzari2012} (Ni-$d$ on the two antiferromagnetic sublattices and O-$p$), constructed from the isolated bands 49--114 (no disentanglement) with 500 iterations from the projections. The Fourier transform of the three $q$-points gives a vertex that resolves the phonon momentum over three layers along $c$ ($R_p=-1,0,+1$); no supercell was computed, and the in-plane $R_p$ are collapsed ($n_{q_1}=n_{q_2}=1$). The $-q$ layer of the vertex was reconstructed from the $+q$ layer by time-reversal symmetry. Polar (Fr\"ohlich) coupling is not included in this vertex (\texttt{lpolar} off).

\emph{Raman and ROA calculations.} The finite-momentum Raman tensors were summed on a $k$-mesh with $n_k=8$ (checked against $n_k=4$ to within 10\%). The Keldysh response at finite phonon frequency used $n_k=8$ and $10$ and the $\Gamma$-point-only vertex (computed on a $1\times1\times1$ $q$-grid and a $4\times4\times4$ $k$-grid), with the Fr\"ohlich coupling added for the LO modes.

\emph{Limitations of the model.} (i) Three layers along $c$ do not fully converge the gradient term: most of its weight lies at distances of 8--14~\AA, which leaves an uncertainty of at least $\sim20\%$, and $q$-grids with six points along $c$ ($\pm\frac{1}{6}$, $\pm\frac{1}{2}$) would be needed. (ii) The Fr\"ohlich coupling enters only the Keldysh calculation: the finite-momentum electron--phonon vertex is computed without polar terms, with TO eigenvectors. For light along $c$ the $E$ modes are TO, so the $E$-line ROA is unaffected, whereas the $A$ modes are LO and are treated with the Fr\"ohlich coupling in the Keldysh calculation only. (iii) Self-consistent Hartree and exchange--correlation local-field corrections are not included; DFPT calculations of natural optical activity show that they can strongly renormalize, and occasionally change the sign of, individual gyrotropic components \cite{ZabaloStengel23}. (iv) The position operator is taken diagonal at the Wannier centres, and intra-Wannier orbital moments and quadrupoles as well as spin--orbit coupling are neglected. The inter-site E1--M1 and E1--E2 contributions, including charge-transfer transitions, are contained in the shifts of the electronic states to $\bar{\bm k}\pm\bm p/2$.


\end{document}